\documentclass{article}
\pdfoutput=1 
\usepackage{amssymb, amsmath,bm,tikz,verbatim}
\usepackage{xcolor,colortbl}
\usepackage{amsthm}
\usepackage{amsfonts}
\usepackage[round]{natbib}
\usepackage[utf8]{inputenc}
\usepackage[T1]{fontenc}
\usepackage{babel}
\usepackage{hyperref}
\usepackage{longtable}
\usepackage{makecell, cellspace, caption}
\usepackage{authblk}
\newcommand{\Y}{\bm{Y}}
\newcommand{\M}{\bm{M}}

\newcommand{\siamesfbi}{SI sub-section 3.1 }
\newcommand{\sinonresponse}{SI section 2 }
\newcommand{\sinonresponsenospace}{SI section 2}
\newcommand{\sisamplingbias}{SI section 1 }

 \usepackage{graphicx}

\usepackage{array}
\newcolumntype{L}[1]{>{\raggedright\let\newline\\\arraybackslash\hspace{0pt}}m{#1}}
\newcolumntype{C}[1]{>{\centering\let\newline\\\arraybackslash\hspace{0pt}}m{#1}}
\newcolumntype{R}[1]{>{\raggedleft\let\newline\\\arraybackslash\hspace{0pt}}m{#1}}
\definecolor{tuftsblue}{RGB}{31,55,108}

\usepackage{titlesec,titletoc,bbm}
\titleformat*{\section}{\normalfont\large\bfseries\color{tuftsblue}}
\titleformat*{\subsection}{\normalfont\large\bfseries\color{tuftsblue}}
\titleformat*{\subsubsection}{\normalfont\bfseries\color{tuftsblue}}
\titleformat*{\paragraph}{\normalfont\bfseries\color{tuftsblue}}
\titleformat{\paragraph}[runin]{\normalsize\bfseries\color{tuftsblue}}{\theparagraph}{1em}{}

\usepackage[noabbrev,capitalize]{cleveref}

\title{Shining a Light on Forensic Black-Box Studies}
\date{}
\author[a,b]{Kori Khan \thanks{corresponding author, kkhan@iastate.edu}}
\author[a,b]{Alicia L. Carriquiry} 

\affil[a]{Department of Statistics, Iowa State University}
\affil[b]{Center for Statistics and Applications in Forensic Evidence (CSAFE), Iowa State University}

\begin{document}

\maketitle

\begin{abstract}
 Forensic science plays a critical role in the United States criminal justice system. For decades, many feature-based fields of forensic science, such as firearm and toolmark identification, developed outside the scientific community's purview. The results of these studies are widely relied on by judges nationwide. However, this reliance is misplaced. Black-box studies to date suffer from inappropriate sampling methods and high rates of missingness. Current black-box studies ignore both problems in arriving at the error rate estimates presented to courts. We explore the impact of each type of limitation using available data from black-box studies and court materials. We show that black-box studies rely on non-representative samples of examiners. Using a case study of a popular ballistics study, we find evidence that these unrepresentative samples may commit fewer errors than the wider population from which they came. We also find evidence that the missingness in black-box studies is non-ignorable. Using data from a recent latent print study, we show that ignoring this missingness likely results in systematic underestimates of error rates. Finally, we offer concrete steps to overcome these limitations.
\end{abstract}

Forensic science is a central part of the United States (U.S.) criminal justice system. However, in the last two decades, it has become apparent that faulty forensic science has caused gross miscarriages of justice \citep{garrett2009invalid,bonventre2021wrongful,fabricant2022junk}. This occurred partly because many forensic disciplines were developed by and for entities outside of the scientific community. A 2009 report commissioned by the  National Academy of Sciences highlighted that, with the exception of DNA, no forensic science field had been empirically shown to be consistent and reliable at connecting a piece of evidence to a particular source or individual \citep{nrc}. This problem was particularly concerning for feature-based comparison methods, such as latent print analysis, firearm and toolmark identification, and footwear impression examinations because these methods are not rooted in science but rather in subjective, visual comparisons. 

\sloppy{
In 2016, a technical report by the President’s Council of Advisors on Science and Technology (PCAST) highlighted that some feature-based comparison methods, like bitemarks, were known to be invalid and were still being used in U.S. courts \citep{pcast}. It also pointed out that there was still no empirical evidence other methods in use were valid. The report stated that empirical testing through ``black-box'' studies is the only scientific way to establish the validity of feature-based comparison methods. In the years that followed, the PCAST report spurred a number of
black-box studies in a variety of fields.}

These studies immediately found their way into the U.S. criminal justice system, at times before peer review. Judges frequently use them when deciding whether or not results from feature-based comparison methods should be admissible. Currently,  all federal courts and the majority of state courts evaluate expert scientific testimony by the \textit{Daubert} standard (or a modified form of it) \citep{Daubert,Kumho,FRE}. Under the \textit{Daubert} standard, a trial judge must assess whether an expert's scientific testimony is based on a scientifically valid methodology that has been properly applied to the facts at issue in the trial. The judge is asked to consider a number of factors. One is whether the theory or technique in question has a ``known or potential’’ error rate. A ``high’’ known or potential error rate would weigh in favor of excluding the testimony at a criminal trial.

Across disciplines, the vast majority of times the ``known or potential’’ error rate is called into question, judges find error rates are low enough to favor admission (e.g., \cite{Cloud}; there are exceptions, see, e.g., \cite{Shipp}). The results of black-box studies are frequently used to support such findings. In relying on the results from black-box studies to evaluate a forensic technique’s admissibility, U.S. judges are explicitly told that the conclusions from the black-box studies can be generalized to results obtained by examiners in the discipline in question. For example, a recent ballistics black-box study asserted that "[This] study was designed to provide a representative estimate of the performance of F/T [firearm and toolmark] examiners who testify to their conclusions in court." \citep{monson2022planning}. The Federal Bureau of Investigation's Lab used this study to support a claim to the Court that "In sum, the studies demonstrate that firearm/toolmark examinations, performed by qualified examiners in accordance with the standard methodology, are reliable and enjoy a very low false positive rate." \citep{FBIResponse2022}.

Unfortunately, these statements are false. Our review of existing black-box studies found that current studies rely on non-representative, self-selected samples of examiners. They also all ignore high rates of missingness, or nonresponse. These flaws, individually and jointly, preclude any statement about discipline-wide error rates. But perhaps more problematically, we also found evidence that, in some cases, these problems work to systematically underestimate the error rates presented to judges. 

\sloppy{
The rest of the paper proceeds as follows. In \cref{sec:blackboxstudies}, we introduce the concept of black-box studies and the accuracy measures they consider. In \cref{samplingbias}, we review the methods used to select examiners for participation in black-box studies. Using a popular ballistics black-box study as an illustrative example, we show that these methods lead to unrepresentative samples of participants. For this case study, we also explore how the methods employed could contribute to lower error rate estimates than may be present in the wider discipline. In \cref{unititemnonresponse}, we explore the extent of the missing data problem in black-box studies.  To put these problems in context, we give a brief overview of the analysis of missing data. Then, in \cref{casestudies}, we use the experimental design and data (to the extent they are available) from two black-box studies. We find evidence that examiners who commit a disproportionate number of errors also have disproportionately high nonresponse rates in black-box studies. Using simulation studies, we show that ignoring this kind of missingness could result in gross underestimates of error rates. We also highlight how misleading the current standards for reporting results in black-box studies are. Finally, in \cref{conclusion}, we offer concrete steps to address these limitations in future studies. }

\section{Black-Box Studies} \label{sec:blackboxstudies}
Forensic feature comparison disciplines are difficult to evaluate empirically. These disciplines, which include latent print analysis, firearm and toolmark examination, and footwear impression examination, rely on inherently subjective methods. In response to this complication, \citet{pcast} stated that the scientific validity of these disciplines could only be evaluated by multiple, independent ``black-box'' studies. 

In a black-box study, researchers accommodate the subjectivity of the method by treating the examiner as a ``black-box.'' No data are collected on how an examiner arrives at his/her conclusions. Instead, the examiner is presented with an item from an unknown origin and an item(s) from a known source and is asked to decide whether the items from known and unknown sources came from the same source. Although the details of arriving at such a decision can vary by discipline, the general steps are the same. First, the examiner assesses the quality of the sample of unknown origin to determine whether it is suitable for comparison. Many disciplines have a categorical outcome for this stage. For example, latent print examiners often use three categories: value for individualization, value for exclusion only, or no value \citep{ulery2011accuracy,eldridge2021testing}. If the item is deemed suitable for comparison, the examiner then arrives at another categorical conclusion about the origin of the unknown. The categories of this conclusion typically include Identification (the samples are from the same source), Exclusion (the samples are from different sources), or Inconclusive (with different disciplines having various special cases of inconclusive). 

\sloppy{
The PCAST report stated that empirical studies must show that a forensic method is accurate, repeatable, and reproducible for the method to have foundational validity \citep{pcast}. Accuracy is measured by the rate at which examiners obtain correct results for samples from the same source and different sources. Repeatability is measured by the rate at which the same examiner arrives at a consistent conclusion when re-examining the same samples. Reproducibility is the rate at which two different examiners reach the same conclusion when evaluating the same sample.
All current black-box studies assess accuracy, and accuracy measures are the most frequently used measures in courts \citep{Shipp,FBIResponse2022,Cloud}. The studies that assess repeatability and reproducibility do so after assessing accuracy. Typically, researchers recruit participants and give a set of items for comparison to assess accuracy. Then, they select (or request volunteers from) a subset of the original participants and distribute new (potentially repeating) items for comparison in the repeatability or reproducibility stage. Because of this setup, the issues we discuss in this paper always apply to repeatability and reproducibility in addition to accuracy. For simplicity, we restrict our attention to accuracy measures.}

Accuracy is typically quantified with four measures: 1) the false positive error rate, 2) the false negative error rate, 3) sensitivity, and 4) specificity. However, the error rates tend to be considered the most important accuracy measures, so we focus on these \citep{smith2016validation}. The false positive error rate focuses on different source comparisons. Researchers typically divide the number of items for which examiners incorrectly concluded "Identification" by the number of total different source comparisons. The false negative error rate focuses instead on the same source comparisons and is defined similarly. 

Importantly, items marked inconclusive are almost always considered ``correct'' comparison decisions. This practice apparently originated in the late 1990s among the feature comparison community. The Collaborative Testing Service (CTS) treated inconclusives as errors until approximately 1998 \citep{CTS_testimony}. The decision to change the treatment of inconclusives was seemingly influenced, in part, by high error rates in ballistics \citep{CTS_testimony}. Treating inconclusives as ``correct'' is problematic \citep[for more details, see][]{inconclusives,dorfman2022inconclusives}. However, there is no agreement on how to handle inconclusives in analyses \citep[see, e.g.,][]{weller2020commentary}. In this paper, we treat inconclusives as ``correct'' decisions. We do so to view the impacts of sampling and nonresponse bias in the light most favorable to the error rates currently being reported to courts across the U.S. 

The PCAST report, and others, have pointed out several desirable features of black-box studies; for example, open set design, blind testing, and independent researchers overseeing the study \citep{pcast}. In the wake of the PCAST report, authors of many black-box studies assume that if they have addressed these elements, they can use their studies' results to make statements about the discipline-wide error rate. However, in this paper, we assume that all the desirable features described in the PCAST report have been met. We show that even for a study where this is the case, the current methods of sampling examiners and handling nonresponse rates preclude any such conclusions. 

\section{Unrepresentative Samples of Examiners} \label{samplingbias}
The inferential goal of current black-box studies is to make a statement about either the discipline-wide error rate or the average examiner's error rate in a specific discipline \citep[see, e.g.,][]{chumbley2021accuracy,smith2021beretta,eldridge2021testing,monson2022planning}. In other words, these studies wish to take observations made on a sample of examiners and arrive at a conclusion about the broader population of examiners to which they belong.

The power of well-done statistics is the ability to do precisely that: take observations made on a sample and generalize these observations to a wider population of interest. However, this ability comes at a price: valid sampling methods must be used to ensure that the sample selected to participate in the study is representative of the larger population. The gold standard for achieving a representative sample is random sampling \citep{levy2013sampling,fisher1992statistical}. In random sampling, members of the population are selected, with known probability, to be included in the study by the researcher. 

Random sampling is desirable for many reasons. For example, it is necessary for many standard statistical techniques. However, most random sampling methods require, at least theoretically, that it be possible to enumerate the population of interest. There are many cases where this is not feasible or practical. The inability to use random sampling does not always preclude researchers from generalizing to the population of interest. \cite[e.g.,][]{smith1983validity,elliott2017inference}. In order to make such generalizations with non-random sampling, however, something must generally be known about the population of interest, and care must be taken to avoid known sources of bias. 

\sloppy{
One such source of bias in non-random sampling methods is selection bias. Selection bias occurs when the sampling method results in samples that systematically over-represent some members of the underlying population. This can result in biased estimates for the parameter of interest, such as the error rate. One of the most well-studied types of selection bias comes in the form of self-selection, or volunteer, bias. This occurs when the researcher does not choose the population members to be in the study but instead allows members of the population to ``volunteer'' to participate. Research in various fields has indicated that this typically results in unrepresentative samples because those who volunteer to participate tend to be different than those who do not volunteer \citep[e.g.,][]{ganguli1998random,dodge2014characteristics,jordan2013volunteer,strassberg1995volunteer,taylor2009examination}. }

\subsection{Sample Selection in Black-Box Studies}
For most feature-comparison disciplines, little is known about the population of examiners analyzing forensic evidence. The standards for serving as an expert witness in a U.S. trial are not high. For example, two days (or less) of formal training in a discipline can be sufficient to qualify as an expert in a forensic science discipline \citep[see, e.g.,][]{Moore}. Few states have established regulatory boards to establish minimum requirements to be an examiner. This, unfortunately, means that it is very difficult to determine what a representative sample of examiners might look like. 

No black-box study attempts to address this information gap. Instead, to our knowledge, every black-box study to date has used self-selected samples of examiners \citep[see, e.g.,][]{eldridge2021testing,baldwin2014study,chumbley2021accuracy,hicklin2021accuracy,smith2016validation,smith2021beretta,ulery2011accuracy,richetelli2020forensic,guttman2022results,hicklin2022accuracy}. These self-selected participants are typically solicited through email listservs for one or more professional organizations. The black-box studies for some disciplines, like latent prints, often accept every volunteer examiner. While this does not alleviate the probable self-selection bias, it ensures that the researchers are not excluding self-selected examiners based on qualities that may be related to error rates. 

On the other hand, black-box studies in other disciplines, such as firearm and toolmark identification, often impose inclusion criteria that reasonably could be expected to be related to error rates. For example, \cite{monson2022planning} reports that the ballistics study we refer to as the FBI/Ames study used self-selected examiners. However, the researchers also restricted participation in the study to ``fully qualified examiners who were currently conducting firearm examinations, were members of [the Association of Firearm and Tool Mark Examiners] AFTE, and were employed in the firearms section of an accredited public crime laboratory within the U.S. or a U.S. territory.''  They also excluded FBI employees to avoid a conflict of interest.

Other than perhaps for the word ``qualified,'' none of these criteria are directly related to a characteristic required of examiners who examine firearm and toolmark evidence or testify about it in U.S. courts. For example, AFTE is a professional organization. To our knowledge, no court has ever deemed membership in AFTE necessary to qualify an examiner as an expert witness. Additionally, many privately employed (or self-employed) examiners are actively testifying.

There has never been an attempt to assess whether inclusion criteria used in black-box studies are representative of examiners. Here, we explore whether the FBI/Ames study criteria would have excluded examiners currently conducting firearm and toolmark examinations for trials. We used Westlaw Edge's collection of expert witness materials to identify 60 unique expert witnesses whose curriculum vitae (CV) indicated the witness was an expert with respect to firearm and toolmark identification (see \sisamplingbias for more details). These CVs cannot be viewed as a representative sample of expert witnesses. Among other problems, Westlaw Edge's materials tend to be disproportionately from federal  jurisdictions. However, they are still useful to explore whether the inclusion criteria used by the FBI/Ames study match the characteristics of expert witnesses who have interacted with courts. 

For each of the 60 expert witnesses, we assessed whether each examiner was a current AFTE member and whether he/she worked for a private or public employer (see \sisamplingbias for more details). In \cref{tab:inclusions_criterion}, we found that just over 60\% of the expert witnesses met each criterion separately, but only 38.3\% met both. In other words, with fewer than two of the inclusion criteria used in the FBI/Ames study, the majority of these expert witnesses would have been excluded from participation.

\begin{center}
\begin{tabular}{| m{5cm} |   m{2cm}  |     }
		\rowcolor{tuftsblue}
		\makecell{\textbf{\textcolor{white}{Criterion}}} & \makecell{\textbf{\textcolor{white}{\% Experts }}}   \\
		\hline
		Current AFTE member &  63.3\%  \\
		\hline
		Public Employer &   65.0\% \\
		\hline
		Current AFTE member \& Public Employer & 38.3\% \\
		\hline
\end{tabular} 	
\captionof{table}{\% of 60 expert witnesses satisfying FBI/Ames Criterion  \label{tab:inclusions_criterion}}
\end{center}

More problematically, some of the inclusion criteria used in black-box studies are either known to be related to or could reasonably be related to error rates. Using the FBI/Ames study as an example again, this study excludes foreign examiners. Yet, foreign examiners can provide testimony in court, and the results of studies on foreign examiners are used to support the admissibility of firearm and toolmark examinations (see \cite{FBIResponse2022} citing \cite{kerkhoff2018part}). However, foreign examiners have also been linked to higher error rates than U.S. examiners in other disciplines \citep[for example, in latent palmar prints, see,][]{eldridge2021testing}. The exclusion of foreign examiners thus should not be done lightly. Additionally, forensic laboratory accreditation requires proficiency testing of examiners and for the lab to assess and certify that the examiners are competent \citep{doyle2020review}. The expectation would be that this process reduces errors. However, examiners are not required to work for an accredited lab to present testimony at trial. Thus, these criteria make it likely that sampling bias is present. In the case of the FBI/Ames study, the particular inclusion criteria used also suggest that this bias works to result in underestimations of error rates.  

\section{Unit and Item Nonresponse in Black-Box Studies} \label{unititemnonresponse}
In this section, we borrow from survey methodology to distinguish between unit and item nonresponse (see, e.g., \citet{little2019statistical} pgs. 5-6 and \citet{elliott2005patterns} pg. 2097). Unit nonresponse occurs when a participant who agreed to participate in a study does not respond to a single assigned comparison. Item nonresponse occurs when a  participant responds to at least one but not all assigned comparisons. The presence of either unit or item nonresponse can lead to bias and a loss of power in statistical analyses which do not account for this missingness \citep{groves2006nonresponse,rubin1976inference}. These problems can be exacerbated in studies, like many black-box studies, where both types of nonresponse are present. Yet, to our knowledge, no one has ever attempted to formally analyze the patterns of missingness in black-box studies or to adjust error rate estimates to account for it. 

\subsection{The Nomenclature of Missing Data}
\sloppy{
To adjust for nonresponse, researchers must first explore the patterns of missingness in their data. Statistical methods to address missing data depend heavily on the mechanisms that caused the missingness to arise. \cite{rubin1976inference} was the first to formalize missing-data mechanisms, but the nomenclature has since changed  \citep{little2019statistical,kim2014statistical}. Currently, missing data mechanisms are described as falling into one of three categories: missing completely at random (MCAR), missing at random (MAR), and not missing at random (NMAR) \citep{little2019statistical,kim2014statistical}. }

To understand the differences in the context of black-box studies, we define $\Y$ to be the $n \times K$ matrix of complete data. This matrix would include, at a minimum, the response for every assigned item. It could also include auxiliary information about the item or responding examiner. We let $\M$  be the $n \times K$ matrix where $M_{ij}$ is $1$ when $Y_{ij}$ is missing and $0$ otherwise. Finally, we let $\bm{\phi}$ be a set of unknown parameters on which the missingness mechanism depends. In the best-case scenario, whether data are missing should not depend on any element of $\Y$. More precisely, the following would hold: 
\begin{eqnarray} \label{eq:MCAR}
f \left( \M | \Y, \bm{\phi} \right) &\equiv & f \left( \M |  \bm{\phi} \right)   ~\forall ~ \Y, \bm{\phi}.
\end{eqnarray} 
In this case, the missingness would be MCAR. If, instead, the missingness depended only on observed values of the data matrix $\Y$, i.e.:
\begin{eqnarray} \label{eq:MAR}
f \left( \M | \Y, \bm{\phi} \right) &\equiv & f \left( \M | \Y_{obs}, \bm{\phi} \right)  ~\forall ~ \Y, \bm{\phi}, 
\end{eqnarray}
then we would call it MAR. The most problematic missingness mechanism is NMAR. This occurs when $f \left( \M | \Y, \bm{\phi} \right)$ cannot be simplified to the right-hand sides of either \eqref{eq:MCAR} or \eqref{eq:MAR}. 

\sloppy{
For many analytical goals, MCAR and MAR do not typically result in biased estimates. Instead, they primarily affect the uncertainty associated with such estimates. In the context of black-box studies, point estimates obtained for error rates may not be systematically skewed from the underlying population error rate. Instead, the confidence intervals associated with these point estimates may need to be adjusted. For sufficiently low nonresponse rates, some researchers consider ignoring MCAR (and sometimes MAR) missingness acceptable. This is part of the reason why MCAR and MAR are often referred to as ignorable missingness. There is no consensus on what ``sufficiently low'' means -- rules of thumb range from 5\% to 10\% \citep{schafer1999multiple,bennett2001can}. Even with ignorable missingness, however, a sufficiently high nonresponse rate can become problematic \citep{madley2019proportion}. Some researchers have proposed that nonresponse rates above 40\% should always preclude generalizations to a wider population \citep{jakobsen2017and,dong2013principled}. }

Unlike MAR and MCAR, NMAR  can lead to bias in the point estimates and distortions in uncertainty estimates even with low rates of nonresponse. It is inappropriate to ignore this type of missingness in statistical analyses.  As a result, NMAR is often referred to as non-ignorable missingness. 

\subsection{Adjusting for Nonresponse} \label{sec:adj}
There are numerous methods to adjust for both unit and item nonresponse. The appropriate method will depend on the type of missingness mechanisms present. Assessing this will typically require a thorough analysis of all collected data. When in doubt about the missingness mechanisms, many researchers recommend conducting sensitivity analyses to assess the potential impacts of different types of mechanisms \citep{pedersen2017missing}.

Most simple methods for handling missing data only result in unbiased estimates if the missing mechanism is MCAR. These methods include complete case analysis and available case analysis. In a complete case analysis, an analysis is only carried out on cases where the full set of analysis variables is observed. An available case analysis, on the other hand, uses all data available about the analysis variables. We emphasize these approaches are only appropriate when the missingness is MCAR \citep{schafer2002missing}. Even in this case, standard errors for the estimators can be adversely impacted. 

Unfortunately, the missingness in real-world data of human subjects is rarely MCAR  \citep[see e.g.,][]{jakobsen2017and,mislevy1996missing}.
When the missingness is not MCAR, the methods of adjusting for missingness almost always require auxiliary information. For MAR approaches, this auxiliary information can be limited to assumptions about the relationship between missingness in one analysis variable and other observed analysis variables. However, methods involving NMAR approaches almost always require auxiliary information beyond the analysis variables \citep{groves2006nonresponse,riddles2016propensity,franks2020nonstandard}. 
\subsection{Implications for Black-Box Studies} \label{subsec:implblackbox}
Black-box studies are plagued by both unit and item nonresponse. We emphasize that we treat inconclusive decisions as observed in this discussion. However, the authors of black-box studies pay little attention to the nonresponse. In fact, they often fail to release enough information to even calculate the relevant nonresponse rates. 

Ideally, it would be possible to calculate the nonresponse rates for each analysis conducted.  For example, to calculate the unit and nonresponse rates for a false positive error rate, we need to know the number of different source items assigned to participants. At the time of writing, only two black-box studies have released sufficient information to calculate item nonresponse rates for both false positive and false negative error rate estimations \citep[see,][]{eldridge2021testing,hicklin2022accuracy}.

In \cref{tab:nonresponse}, we provide unit and item nonresponse rates for black-box studies aggregated over all potential analyses. This table is limited to black-box studies that released sufficient information to calculate both unit and item nonresponse rates (see \sinonresponsenospace). These unit nonresponse rates reflect those seen in other black-box studies that do not release sufficient information to calculate item nonresponse (see \sinonresponse for more details about these and other studies). We note that unit nonresponse rates are often over 30\%, far above any definition of ``low'' nonresponse rates \citep[see e.g.,][]{ulery2011accuracy,eldridge2021testing,richetelli2020forensic,smith2021beretta}.

\begin{center}
\begin{tabular}{| L{7cm} |   L{1.5cm}|  L{1.5cm}|   }
		\rowcolor{tuftsblue}
	\small	\makecell{\textbf{\textcolor{white}{Study}}} & \small \makecell{\textbf{\textcolor{white}{Unit}} } & 
	\small	\makecell{\textbf{\textcolor{white}{Item }}}  \\ 
		\hline
\cite{ulery2011accuracy} (Prints) & 33.5\% & 0.2\% \\
		\hline
		\cite{baldwin2014study} (Cartridges) & 23.2\% & .06\% \\ 
		\hline 
		FBI/Ames Study (Bullets)  & $\geq$ 32.4\% &  35.6\%  \\
		\hline
		FBI/Ames Study (Cartridges)  & $\geq$ 32.4\% &  35.1\%  \\
		\hline
		\cite{eldridge2021testing} (Palmar Prints) & 31.1\% & 27.5\% \\
		\hline
	\cite{richetelli2020forensic} (Footwear Impressions) & 33.0\% &  9.6\% \\ 
		\hline
\end{tabular}
\captionof{table}{Example of nonresponse rates in black-box studies  \label{tab:nonresponse} }
\end{center}

In \cref{tab:nonresponse}, we have calculated nonresponse rates assuming that any recorded response is observed, even if no comparison decision was rendered. However, practically speaking, only a recorded comparison decision can be used to estimate error rates. If only comparisons with a comparison decision are considered observed, the item nonresponse could be much higher than those given in \cref{tab:nonresponse}. 

To illustrate this, we consider one of the two black-box studies that have released sufficient information to calculate item nonresponse rates for the false positive rate. In \cite{eldridge2021testing}, 328 participants enrolled in the study, and 226 actively participated. Each participant was assigned 22 different source comparisons. Twenty-five of the 226 active participants failed to start all 22 different source comparisons assigned to them. Another 3 participants stated the image was of no value for every different source comparison they responded to. Finally, another 13 failed to render a single comparison decision for their assigned different source comparisons, despite finding at least one such comparison to be suitable for comparison. Out of the original 328 enrolled participants, only 185 rendered a comparison decision on at least one different source comparison. For the purpose of calculating the false positive error rate, there was a unit nonresponse rate of $ \left( 1- \frac{185}{328}\right) \times 100 = 44$\%. The 185 participating examiners rendered 2,560 comparison decisions for their assigned $185 \times 22 = 4,070$ different source comparisons. The item nonresponse rate was thus $ \left( 1 - \frac{2560}{4070} \right) \times 100 = 37$\% for the false positive error rate calculations in this study (where we treat inconclusives as observed). As we have discussed, such high rates of missingness preclude generalization to a broader population of examiners. 

Despite the high nonresponse rates, authors of black-box studies have yet to report examining the patterns of missingness in their data. Black-box studies do not attempt to adjust for the missingness in their analyses. Instead, there are two ways that black-box studies deal with missing data. Some black-box studies drop a participant from the analysis if the participant did not answer all items assigned to him/her \citep{ulery2011accuracy,hicklin2021accuracy}. This is an example of complete case analysis discussed in \cref{sec:adj}, and it is only appropriate for MCAR missingness. The second way black-box studies handle missingness is to analyze the observed responses, ignoring the nonresponse. This approach is marginally better than dropping participants with missing items and is most similar to an available cases approach. However, again this is only potentially defensible if the missingness is MCAR. 

It is highly unlikely that the missingness in black-box settings is MCAR. Technically, MCAR missingness would require a random sample (as opposed to self-selected samples) of examiners. Outside of forensic science, missingness is typically assumed to be potentially non-ignorable in testing settings \citep{mislevy1996missing,pohl2014dealing,dai2021handling}.  In such settings, \citet{mislevy1996missing} suggests "intuition and empirical evidence" support that "[E]xaminees are more likely to omit items when they think their answers are incorrect than items they think their answer would be correct." If an examinee is proficient enough to know when he/she is likely to be incorrect, then this type of behavior will lead to an underestimate of error rates if missingness is ignored. 

To appropriately adjust for missingness in black-box studies, researchers will likely need auxiliary information. This information could come in the form of examiner characteristics or item characteristics. Many black-box studies collect this type of information, but only one has released any portion of such data to the public in a meaningful way. Indeed, most black-box studies fail to release any de-identified data at all. Instead, as we will explore in \cref{amesfbicasestudy}, they give aggregated summaries that could be misleading in the context of high nonresponse rates.

The authors of black-box studies typically reject the possibility of non-ignorable missingness. Many state that missingness occurs because examiners are too busy to participate in the study or, alternatively, to complete all items if they do participate \citetext{Private communication with Heidi Eldridge, 2021}. However, it is possible for missingness to be non-ignorable and for nonresponse to be due to examiners being too busy. For example, busy examiners may choose to respond to the easier item comparisons as they take less time. Rather than speculating, however, the appropriate step would be to assess the patterns of missingness in the data. As no one has done this, we offer the first such attempt to do so now.

\section{Case Studies} \label{casestudies}
In this section, we focus on the potential impact of item nonresponse for error rate estimates in black-box studies. As previously referenced, only two black-box studies have released data in a form that an independent researcher can analyze (i.e., the data are released with enough detail about the study design that, at a minimum, the item nonresponse rates can be calculated for individual analyses). In \cref{eldridgecasestudy}, we use one of these datasets to explore some of the patterns of item nonresponse. We show there is evidence of non-ignorable missingness. Using that insight, we use simulation studies to replicate the FBI/Ames study's exploration of false positive error rates in bullet comparisons in \cref{amesfbicasestudy} and highlight how misleading the current trends of reporting responses in black-box studies can be. 

\subsection{EDC Study: Palmar Prints} \label{eldridgecasestudy}
\sloppy{
This sub-section focuses on the item nonresponse in a study described in \cite{eldridge2021testing} (hereafter the EDC study). This study assessed the accuracy of latent print examiners' analysis of palmar prints. Each participant was asked questions about his/her demographic information, training, and employer. All participants then received 75 items (comparisons) to complete. 
The study design assumed participants in this study followed a multi-stage approach to analyzing items. First, examiners were asked to assess the images of the prints for suitability for comparisons. If examiners found an item suitable for comparison, they could enter a conclusion of Inconclusive, Identification, or Exclusion. As part of their analysis, examiners were also asked to rate each item's comparison difficulty. In this section, we treat any response to an item as non-missing. Thus, we consider items marked as not suitable for comparisons as observed.} 

The authors released information about all items that examiners responded to and demographic information for most participants. They also released fairly rich information about the quality of images for compared items. When asked, they declined to release information about the comparisons examiners did not respond to. 

We begin the analysis of item nonresponse with variables that have been previously linked to non-ignorable missingness. As alluded to in \cref{subsec:implblackbox}, one common pattern of non-ignorable missingness on assessment tests is when examinees fail to respond to items they believe they would answer incorrectly. In this dataset, examiners were asked to rate each item's difficulty on a Likert-type scale, including possible responses of: Very easy/obvious, Easy, Moderate, Difficult, and Very difficult. For the items examiners responded to, 910 were deemed Very easy/obvious, and only 545 were deemed  Very difficult. In seven cases, examiners ranked the item difficulty level and then failed to give a comparison decision. These items were all marked as either Moderate or Very Difficult. These patterns suggest that examiners were more likely to respond to items they deemed Very easy and less likely to respond to items they deemed Very difficult. As \cite{eldridge2021testing} (and others outside of black-box studies) have observed, more errors are committed on items ranked as difficult. Thus, there is evidence of non-ignorable missingness here. Ideally, we would have a list of every item assigned to each examiner. While it is not possible to know how a particular examiner would have ranked a particular item, it would be possible to use auxiliary information (e.g., other examiner's rankings or information about the quality of items) to more formally assess whether the items with no response were more likely to be viewed as difficult.  

There are other ways to use the released data to formally assess whether there is evidence of non-ignorable missingness. Because we know that each examiner was assigned 75 items, we can calculate each examiner's item nonresponse rate. The study's authors also identified various examiner characteristics they claimed were associated with higher error rates. If the item nonresponse was ignorable, we should not see any relationship between high rates of item nonresponse and characteristics associated with high error rates. We can use permutation tests to formally assess whether a statistically significant relationship exists between high degrees of item nonresponse and examiner characteristics associated with high error rates.

\sloppy{
To do this, we restrict our attention to the 197 examiners who released both demographic information and at least one response. We define an examiner to have a high degree of item nonresponse if he/she failed to respond to over half of the 75 assigned items. The EDC study identified several characteristics that were associated with high error rates. For example, participants employed outside of the United States made half of the false positive errors in the study despite only accounting for 18.1\% of the 226 active study participants. Similarly, the EDC study noted that non-active latent print examiners (LPEs) disproportionately made false positive error rates. Using machine learning approaches, the EDC study also noted that working for an unaccredited lab and not completing a formal training program were weakly associated with higher error rates (lower accuracy) among examiners. We note that these last two observations relied on analyses that, themselves, could have been impacted by missing data. However, we take these findings at face value here.} 

For the permutation tests, we focus on the four characteristics associated with higher error rates: working for a non-US employer, being a non-active LPE, working for an unaccredited lab, and not completing a formal training program. Because each of the four characteristics under consideration is binary, we can use the same general approach. We explain the methodology by focusing on whether examiners work for a non-US employer. 

We consider the following hypothesis test:

\begin{itemize}
  \item[] \footnotesize $H_0$:  Foreign-employed examiners and U.S.-employed examiners are equally likely to leave over 50\% of their items blank.
  \item[]\footnotesize $H_A$: Foreign-employed examiners are more likely than U.S.-employed examiners to leave over 50\% of their items blank.
\end{itemize}
Thirty-eight (19.2\%) of the 197 examiners worked for non-U.S. employers. Under the null hypothesis, we would expect that approximately 19.2\% of the examiners with high rates of missingness worked for non-U.S. employers. Instead, we observe that 28.6\% (or 14 examiners) of the 49 examiners with a high degree of item nonresponse worked for non-U.S. agencies.

If the null hypothesis were true, the probability that 28.6\% or more of the examiners with a high degree of item nonresponse worked for foreign agencies would be about 4.9\%. We refer to this probability as the p-value of the hypothesis test specified above. Because it is low, there is weak evidence that examiners working for non-U.S. agencies are not only more error-prone, but they are also more likely to fail to respond to over half of their assigned items. Note, the terminology ``weak evidence'' comes from historical hypothesis testing, where the decision to fail to reject or reject a null hypothesis was often made based on whether a p-value was less than .05. This type of decision-making can be problematic \citep{wasserstein2016asa}. Here, we emphasize that we view this p-value as more of a data exploration tool: the closer it gets to zero, the less reasonable it is to assume that the item nonresponse is operating independently of the considered characteristic (in this case, type of employer). Here, we believe a p-value of 4.9\% warrants further investigation before using analyses that ignore missingness.

We can repeat this procedure for the remaining three characteristics identified by \cite{eldridge2021testing} as associated with high error rates. When we do so, the p-values for the corresponding hypothesis tests involving being a non-active LPE, working for a non-accredited lab, and having not completed a formal training program are 46.6\%, 3.1\%, and 44.7\%, respectively. Thus, there is evidence that examiners who work for an unaccredited lab are more likely to fail to respond to over 50\% of the items assigned to them than examiners who work for an accredited lab. However, there is no evidence of such a relationship for either being a non-active LPE or having not completed a formal training program. In sum, two of the four characteristics identified by \cite{eldridge2021testing} as being associated with higher error rates also may be associated with higher rates of item nonresponse. 

To summarize, there is evidence that examiners are more likely to respond to Very easy items than Very difficult items. Furthermore, some of the more error-prone examiners also are more likely to leave over half of their items blank than their less error-prone counterparts. These trends are evidence of non-ignorable missingness. The association between high nonresponse and higher error rates suggests that ignoring this missingness will result in underestimates of the associated error rates. However, because there is no information about items each examiner was assigned but chose not to respond to, it's difficult to develop a formal attempt to adjust for missingness in error rate calculations. Thus, in the next section, we use simulation studies to demonstrate the potential impact of non-ignorable missingness. 

\subsection{FBI/Ames Study: Bullet Comparisons} \label{amesfbicasestudy}
In this sub-section, we use simulation studies to demonstrate the impact that item nonresponse can have on estimates of error rates in black-box studies. We do not claim that any of the results are an indication of the truth of a particular existing black-box study. The EDC study suggests that missingness likely depends on the characteristics of both examiners and comparisons. There are no released data to account for these things in a formal statistical way. Instead, our primary purpose is to use a simple model to illustrate how misleading the current methods of reporting error and nonresponse rates can be. We base our simulations on the FBI/Ames black-box study previously considered in \cref{samplingbias}. 

The FBI/Ames study was conducted by researchers at the Ames Laboratory and the Federal Bureau of Investigation (FBI). The study's purpose was to assess the performance of forensic examiners in firearm and toolmark identifications \citep{chumbley2021accuracy,monson2022planning}. It aimed to assess accuracy, repeatability, and reproducibility for both bullet and cartridge comparisons. Although the statistical analyses of collected data remained unpublished until October of 2022, the results were (and continue to be) used to support the admissibility of ballistics expert testimony in criminal trials prior to peer review \citep{Shipp}. Importantly, the repeatability and reproducibility analyses remain unpublished, but continue to be used in courts nationwide \citep[note, these analyses have been challenged by others, see, e.g.,][]{dorfman2022re}. Initially, the FBI/Ames study  released no data capable of being independently analyzed. When we requested the de-identified data to explore patterns of missingness, an Ames lab researcher stated the FBI had not given Ames lab researchers permission to share such data. Our requests for clarification about missingness in these studies remain unanswered. We note that the authors subsequently released some data from the accuracy stage in \citet{monson2023accuracy} while this paper was under review; however, these data were limited to only the observed responses for the accuracy stage. The reporting methods of the FBI/Ames study prior to this partial release of data remain the predominant practice in the field. Because our simulation studies are meant to highlight how misleading such reporting methods can be and are not meant to be a reflection of the truth of the FBI/Ames study, we first focus on the methods used prior to the publication of \citet{monson2023accuracy}. As only results from the accuracy stage are publicly available, we focus on accuracy measures.  For simplicity, we further limit our review to the false positive error rates for bullet comparisons. 

The FBI/Ames study reported an estimated false positive error rate of approximately $.7\%$. This estimate, like all estimates in black-box studies, did not account for missingness.  Instead, it was based on the 20 observed false positive errors made in 2,842 observed comparison decisions. We note that 2,891 decisions for the accuracy stage were recorded, but 49 of the responses were dropped from the analysis by the original authors (see \cite{chumbley2021accuracy}). 

We estimate the item nonresponse rate for bullet comparisons across all stages of the study is approximately 35\%. Our estimate assumes the FBI/Ames study authors gave the 173 self-selected examiners 6 packets (see \siamesfbi for details on this estimate). The FBI/Ames study authors report that each packet contained  15 bullet comparisons, across all stages of the study. There is no published paper reporting the number of packets assigned in the accuracy stage, but the abstract of an unpublished paper \citep{bajic2020validation} reports each examiner was intended to receive 2 packets in the accuracy stage (see \siamesfbi for more details; note that the partially released data in \citet{monson2023accuracy} indicate at least one examiner was assigned more than 2 packets in the accuracy stage). Thus, published papers do not provide sufficient information to calculate the item nonresponse rate for the accuracy stage of the study generally. To our knowledge, no papers, published or unpublished, currently report the number of different source items assigned in the accuracy stage, which makes it impossible to calculate the item nonresponse for the false positive accuracy error rate specifically.

If missingness is non-ignorable, as the percentage of missing items increases, the bias of the estimate obtained from an analysis that does not account for missingness will increase. To view the impact of the missingness in the light most favorable to the current estimates, we attempt to use the study design to make a reasonable estimate of the item nonresponse for assigned different source comparisons. Specifically, we assume that 2 packets were assigned to each examiner in the accuracy stage. The FBI/Ames study reports that approximately $2/3$rds of all items were different source comparisons. Based on the number of recorded responses for different source comparisons and these assumptions, the item nonresponse for different source bullet comparisons was at least 17.9\% (see \siamesfbi for more details on the steps of this estimation).

We design a set of simulation studies to mirror the described experiment. To simulate our data, we let $Y_{ij}$ be an indicator of whether examiner $i$ makes an error on item $j$. We define $M_{ij}$ to be an indicator of whether examiner $i$'s response to item $j$ is missing.  We assume that there are $173$ participants who are each given $20$ comparison items. We generate the data in the following way: 
\begin{eqnarray*}
	P(Y_{ij} = 1) \overset{ind}{\sim} \textrm{Bern} \left(p_i \right), \\ 
	P(M_{ij}= 1 | Y_{ij} = 1 ) \overset{ind}{\sim} \textrm{Bern} \left( \pi_i  \right)  \\ 
	P(M_{ij}  = 1 | Y_{ij} = 0 ) ) \overset{ind}{\sim} \textrm{Bern} \left( \theta_i \right),
\end{eqnarray*}   
for $i = 1, \ldots, 173, j= 1, \ldots, 20$. We note that one way missingness could be ignorable is if $\pi_i = \theta_i$ for all $i$. However, given our results in the EDC case study, we explore how a range of non-ignorable missingness mechanisms could impact inference. Throughout all possible scenarios, we ensure that there is always approximately 17.9\% item nonresponse.

The parameter $\pi_i$ represents the probability that an examiner fails to respond to an item on which he/she would have made an error. An examiner is more likely to have a missing response for an item they would have made an error on as $\pi_i$ increases in value. We let $\pi_i = \pi$ be constant across individuals. We note that this is likely not the case in real black-box studies: The EDC data suggest that examiners with different error rates respond at different rates (i.e., likely have different $\pi_i$s). However, this simulation study is meant to give a simple illustration of the wide range of possibilities in observed datasets when missingness is ignored, so we proceed with this simple model. We consider 101 potential values for $\pi$, varying between $0$ and $1$. For each value of $\pi$, we simulate responses for all 173 examiners on 20 items. 

In \cite{chumbley2021accuracy} the authors stressed there was evidence error rates were not constant across examiners. They also noted that 10 of the 173 examiners committed all 20 of the false positive errors. To mirror this, $p_i$ is randomly generated from a uniform distribution on $[0,.007]$ for the first 163 examiners. For the other 10 examiners, $p_i$ is randomly generated from a uniform distribution on $[.55,.6]$. In this way, each examiner has his/her own error rate, and we also reflect the pattern observed in the FBI/Ames study of having 10 examiners more error-prone than the others. To ensure that approximately 17.9\% of the items are missing, $\theta_i$ is chosen as a function of $\pi$ and $p_i$ (see \siamesfbi for more details).

For each value of $\pi$, we really have two datasets: the ``observed'' dataset (restricted to cases where $M_{ij}=0$) and the ``full'' dataset. We calculate the false positive error rate estimate and the 95\% confidence interval for both the observed data and the full data. The FBI/Ames study used a beta-binomial model to arrive at the error rate estimates and 95\% confidence intervals reported in \cite{chumbley2021accuracy}. The authors state the estimates were produced with R packages including \texttt{VGAM} \citep{yee1996,yee2010vgam,yee2015book}. The beta-binomial model in \texttt{VGAM} is numerically unstable, as the expected information matrices can often fail to be positive definite over some regions of the parameter space \citep{yee2022package}. This makes it unsuitable for simulation studies. Instead, we use the Clopper-Pearson estimator \citep{clopper1934use}. In settings like this simulation study, the primary difference will typically be in the width of the confidence intervals- with the beta-binomial model expected to be wider. For example, the Clopper-Pearson estimate and confidence interval for the FBI/Ames false positive error rate are .7\% and (.4 \%,1.1\%), while the corresponding estimates from the beta-binomial model are .7\% and (.3\%,1.4\%).  

It is possible to grossly underestimate the false positive error rate by ignoring the 17.9\% nonresponse rate. For example, in these simulations, the observed error rate estimate is 0\%, and the full error rate estimate is 3.8\% when $\pi=1$. More generally, the  false positive error rate estimate for the full data tends to be between 3\% and 4\% across all values of $\pi$ (see \siamesfbi for more details).  On the other hand, the estimate of the false positive rate for the observed data ranges from 0\% (when $\pi =1$) to over 4.5\% (when $\pi =0$). 

With sufficient data, it may be possible to rule out some of the more extreme discrepancies between the estimates obtained for the observed data and the full data. However, black-box studies do not report such details. For example, before \citet{monson2023accuracy}, the FBI/Ames study only reported the number of examiners with 0 false positives, 1 false positive, and 2 or more false positives, similar to \cref{tab:summaryfalsepos} (they also report the equivalent for false negatives) \citep{bajic2020validation}. We calculated the same type of summary for each of our observed simulated datasets. As shown in \cref{tab:summaryfalsepos}, we obtained summary statistics equivalent to those reported in the FBI/Ames Study when $\pi=.87$. 

\begin{center}
\begin{tabular}{| L{4cm} |   L{1.33cm}|  L{1.33cm} | L{1.33cm}| }
		\hline
		\rowcolor{tuftsblue}
		\makecell{\textbf{\textcolor{white}{Data}}} & \makecell{\textbf{\textcolor{white}{No FP}} } & 
		\makecell{\textbf{\textcolor{white}{1 FP }} } &
		\makecell{\textbf{\textcolor{white}{2 $\color{white} +$ FP}} } \\
		\hline
		FBI/Ames (Obs. Data) & 163 &  5 & 5 \\
	\hline
	Simulated (Obs. Data) & 163 & 5 & 5 \\
		\hline
\end{tabular}
\captionof{table}{False Positives by Examiners}
\label{tab:summaryfalsepos}
\end{center}

For this observed dataset, just like in the FBI/Ames study, there are 20 errors made by only 13 examiners. The Clopper-Pearson estimate and 95\% confidence intervals were equivalent to those for the FBI/Ames study, as shown in \cref{tabl:falseposest}. However, if the error rate had been calculated on the corresponding full dataset, it would have been 3.6\% instead of .7\%. In other words, the ``true'' false positive error rate would have been over 414\% greater than the reported one.

\begin{center}
\begin{tabular}{| L{4cm} |   L{1.5cm}|  L{2.5cm}|  }
	\hline
	\rowcolor{tuftsblue}
\small	\makecell{\textbf{\textcolor{white}{Data}}} & \small \makecell{\textbf{\textcolor{white}{FP Rate}} } & 
\small	\makecell{\textbf{\textcolor{white}{95\% CI }} } \\
	\hline
	FBI/Ames (Obs. Data) & 0.7\% & (0.4\%, 1.1\%)    \\
	\hline
	Simulated (Obs. Data) & 0.7\% & (0.4\%, 1.1\%)    \\
	\hline
	Simulated (Full) & 3.6\% & (3.0\%, 4.3\%) \\ 
	\hline
	FBI/Ames (Full) & ? & ? \\
	\hline
\end{tabular}
\captionof{table}{Clopper-Pearson Error Rates and CI for Observed and Full Datasets}
\label{tabl:falseposest}
\end{center}

We emphasize here that even if the item nonresponse rate was different than 17.9\% or the missingness mechanism is not similar to the one explored in our simulation studies, the general principles from these simulation studies would hold.

\sloppy{
We now take a moment to compare these simulation studies to the partially released information for the accuracy stage in \citet{monson2023accuracy}. We applaud the authors of the FBI/Ames study for releasing some information, but we note the information released is still inadequate for meaningful exploration of the missingness. To explore missingness, the unobserved is as important as the observed. The data released with \citet{monson2023accuracy} included only the observed responses for examiners (see \siamesfbi for more information), and no further information was released about the assigned items that did not receive a response. We note the information released was sufficient to illustrate that the simple simulation studies are not representative of the missingness patterns observed in the data. In the observed data, all false positives were committed by examiners with a 0\% nonresponse rate. Our simulation study included examiners with false positives with non-zero item nonresponse. However, the released data still do not allow an explicit calculation of the item nonresponse for false positive rates (or false negative rates).  To allow others to assess the potential impact on nonresponse, nonresponse rates must be explicitly reported (or sufficient information about the data and study design must be reported so these values can be calculated). To adjust for nonresponse, researchers need, at minimum, the assigned items for each examiner, the examiner's response (or lack of response) to each assigned item, a way to link responses to unique items across examiners, and a way to link examiner demographics to item responses \citep[for more details on the information needed to adjust for missingness, see,][]{khan2023hierarchical}.} We note that the nonresponse rate in the released data was either 50\% or 0\% for each examiner. In such a situation, it is critical to examine the demographic differences between low responders and complete responders and the characteristics of items with no response.


\section{Discussion}  \label{conclusion}

Two major issues currently affect all black-box studies: self-selected participants and large proportions of missingness that go unaccounted for in the statistical analyses of examiner responses. We are the first to explore either of these issues in black-box studies. Using real-world court materials, we have shown that black-box studies are likely relying on unrepresentative samples of examiners. Similarly, we have used actual black-box data to show the missingness in forensic black-box studies is likely non-ignorable. Current estimates of error rates could be significantly biased, and we show there is evidence this bias works to underestimate error rates.

There are ways to overcome both of these problems. The nonresponse rates are easier to address. There is a rich literature on methods to identify missing data mechanisms, adjust statistical analyses to minimize nonresponse bias, and properly account for the associated estimation uncertainty. It would be relatively simple to produce less biased, more reliable estimates of error rates in black-box studies, even without collecting additional data. To do this, however, authors of black-box studies must share enough de-identified information about the participants and the experimental design to enable independent researchers to conduct their own explorations. This information must include the items assigned to each examiner for each analysis and the associated responses (including a nonresponse). When demographic information is available, this information must be linkable to the response data. Similarly, when multiple examiners evaluate the same comparison, their responses must also be linkable. These practices are standard in other scientific disciplines but rare in forensic science.

More difficult to address is the question of the representativeness of black-box study participants. As a threshold matter, participants should not be allowed to self-select. Most black-box studies currently use members of professional organizations, and there is no reason that random samples cannot be taken from these lists to invite examiners to participate in black-box studies. While this would not ensure representative samples, it would at least give more insight into unit nonresponse from a broader population. Beyond this threshold issue, too little is known about individuals who examine forensic evidence. A huge challenge is that almost anyone can be admitted as an expert by a judge. Even identifying members of the relevant population is an elusive problem.  The lack of accessible data collected by courts at all levels adds to the challenge. In the short term, this means that further research about the population of people examining forensic evidence is required. As courts continue to transition to electronic court filings, there will be more opportunities to explore court records. For example, one plausible study to identify examiners could involve a multi-stage sampling design: (1) First, draw a random sample of courts that is itself representative; (2) In each, enumerate criminal cases filed in the last year; (3) Identify experts who testified in these (or a random subset of these) cases and request curriculum vitae from the respective representation. We have completed a pilot study in a single-state jurisdiction similar to this setup and shown that it is possible (albeit difficult) to identify testifying experts in this way. At the moment, obtaining the curriculum vitae of experts who were not subject to an objection relies on the cooperation of state and defense attorneys. 

In this paper, we have focused on how black-box studies are currently being used in courts. In courts, judges are told that these studies can be considered representative of a broader population of studies, and this has shaped our emphasis on representative samples. However, we acknowledge that judges (and juries) may be interested in simply assessing how much weight to give an examiner's testimony rather than determining its admissibility. In other words, they may want to know which type of training or experience can help to improve an examiner's accuracy. In this case, estimating an error rate for all examiners in a given discipline is not the inferential goal. However, from a practical standpoint, research into the population of people examining forensic evidence will still be necessary to begin to understand the factors that may influence an examiner's credibility.

\newpage

\textbf{Acknowledgements}
This work was partially funded by the Center for Statistics and Applications in Forensic Evidence (CSAFE) through Cooperative Agreements 70NANB15H176 and 70NANB20H019 between NIST and Iowa State University, which includes activities carried out at Carnegie Mellon University, Duke University, University of California Irvine, University of Virginia, West Virginia University, University of Pennsylvania, Swarthmore College and University of Nebraska, Lincoln.

\bibliographystyle{plainnat}
\bibliography{references}

\begin{thebibliography}{65}
\providecommand{\natexlab}[1]{#1}
\providecommand{\url}[1]{\texttt{#1}}
\expandafter\ifx\csname urlstyle\endcsname\relax
  \providecommand{\doi}[1]{doi: #1}\else
  \providecommand{\doi}{doi: \begingroup \urlstyle{rm}\Url}\fi

\bibitem[Bajic et~al.(2020)Bajic, Chumbley, Morris, and
  Zamzow]{bajic2020validation}
Stanley Bajic, L~Scott Chumbley, Max Morris, and Daniel Zamzow.
\newblock Validation study of the accuracy, repeatability, and reproducibility
  of firearm comparisons.
\newblock Technical report, Ames Lab., Ames, IA (United States), 2020.

\bibitem[Baldwin et~al.(2014)Baldwin, Bajic, Morris, and
  Zamzow]{baldwin2014study}
David~P Baldwin, Stanley~J Bajic, Max Morris, and Daniel Zamzow.
\newblock A study of false-positive and false-negative error rates in cartridge
  case comparisons.
\newblock Technical report, AMES LAB IA, 2014.

\bibitem[Bennett(2001)]{bennett2001can}
Derrick~A Bennett.
\newblock How can i deal with missing data in my study?
\newblock \emph{Australian and New Zealand Journal of Public Health},
  25\penalty0 (5):\penalty0 464--469, 2001.

\bibitem[Bonventre(2021)]{bonventre2021wrongful}
Catherine~L Bonventre.
\newblock Wrongful convictions and forensic science.
\newblock \emph{Wiley Interdisciplinary Reviews: Forensic Science}, page e1406,
  2021.

\bibitem[Chumbley et~al.(2021)Chumbley, Morris, Bajic, Zamzow, Smith, Monson,
  and Peters]{chumbley2021accuracy}
L~Scott Chumbley, Max~D Morris, Stanley~J Bajic, Daniel Zamzow, Erich Smith,
  Keith Monson, and Gene Peters.
\newblock Accuracy, repeatability, and reproducibility of firearm comparisons
  part 1: Accuracy.
\newblock \emph{arXiv preprint arXiv:2108.04030}, 2021.

\bibitem[Clopper and Pearson(1934)]{clopper1934use}
Charles~J Clopper and Egon~S Pearson.
\newblock The use of confidence or fiducial limits illustrated in the case of
  the binomial.
\newblock \emph{Biometrika}, 26\penalty0 (4):\penalty0 404--413, 1934.

\bibitem[Dai(2021)]{dai2021handling}
Shenghai Dai.
\newblock Handling missing responses in psychometrics: Methods and software.
\newblock \emph{Psych}, 3\penalty0 (4):\penalty0 673--693, 2021.

\bibitem[{Daubert v. Merrell Dow Pharms., Inc.}()]{Daubert}
{Daubert v. Merrell Dow Pharms., Inc.}
\newblock {509 U.S. 579 (1993)}.

\bibitem[Dodge et~al.(2014)Dodge, Katsumata, Zhu, Mattek, Bowman, Gregor, Wild,
  and Kaye]{dodge2014characteristics}
Hiroko~H Dodge, Yuriko Katsumata, Jian Zhu, Nora Mattek, Molly Bowman, Mattie
  Gregor, Katherine Wild, and Jeffrey~A Kaye.
\newblock Characteristics associated with willingness to participate in a
  randomized controlled behavioral clinical trial using home-based personal
  computers and a webcam.
\newblock \emph{Trials}, 15\penalty0 (1):\penalty0 1--7, 2014.

\bibitem[Dong and Peng(2013)]{dong2013principled}
Yiran Dong and Chao-Ying~Joanne Peng.
\newblock Principled missing data methods for researchers.
\newblock \emph{SpringerPlus}, 2\penalty0 (1):\penalty0 1--17, 2013.

\bibitem[Dorfman and Valliant(2022{\natexlab{a}})]{dorfman2022inconclusives}
Alan~H Dorfman and Richard Valliant.
\newblock Inconclusives, errors, and error rates in forensic firearms analysis:
  Three statistical perspectives.
\newblock \emph{Forensic Science International: Synergy}, page 100273,
  2022{\natexlab{a}}.

\bibitem[Dorfman and Valliant(2022{\natexlab{b}})]{dorfman2022re}
Alan~H Dorfman and Richard Valliant.
\newblock A re-analysis of repeatability and reproducibility in the
  {Ames-USDOE-FBI} study.
\newblock \emph{Statistics and Public Policy}, 9\penalty0 (1):\penalty0
  175--184, 2022{\natexlab{b}}.

\bibitem[Doyle(2020)]{doyle2020review}
Sean Doyle.
\newblock A review of the current quality standards framework supporting
  forensic science: Risks and opportunities.
\newblock \emph{Wiley Interdisciplinary Reviews: Forensic Science}, 2\penalty0
  (3):\penalty0 e1365, 2020.

\bibitem[Eldridge et~al.(2021)Eldridge, De~Donno, and
  Champod]{eldridge2021testing}
Heidi Eldridge, Marco De~Donno, and Christophe Champod.
\newblock Testing the accuracy and reliability of palmar friction ridge
  comparisons--a black box study.
\newblock \emph{Forensic Science International}, 318:\penalty0 110457, 2021.

\bibitem[Elliott et~al.(2005)Elliott, Edwards, Angeles, Hambarsoomians, and
  Hays]{elliott2005patterns}
Marc~N Elliott, Carol Edwards, January Angeles, Katrin Hambarsoomians, and
  Ron~D Hays.
\newblock Patterns of unit and item nonresponse in the cahps{\textregistered}
  hospital survey.
\newblock \emph{Health services research}, 40\penalty0 (6p2):\penalty0
  2096--2119, 2005.

\bibitem[Elliott and Valliant(2017)]{elliott2017inference}
Michael~R Elliott and Richard Valliant.
\newblock Inference for nonprobability samples.
\newblock \emph{Statistical Science}, 32\penalty0 (2):\penalty0 249--264, 2017.

\bibitem[Fabricant(2022)]{fabricant2022junk}
M~Chris Fabricant.
\newblock \emph{Junk Science and the American Criminal Justice System}.
\newblock Akashic Books, 2022.

\bibitem[{Fed. R. Evid. 702}()]{FRE}
{Fed. R. Evid. 702}.

\bibitem[{Federal Bureau of Investigation}()]{FBIResponse2022}
{Federal Bureau of Investigation}.
\newblock {FBI Laboratory Response to the Declaration Regarding Firearms and
  Toolmark Error Rates Filed in Illinois v. Winfield}, aff. filed in illinois
  v. winfield dated {May 3, 2022}.

\bibitem[Fisher(1992)]{fisher1992statistical}
Ronald~Aylmer Fisher.
\newblock Statistical methods for research workers.
\newblock In \emph{Breakthroughs in statistics}, pages 66--70. Springer, 1992.

\bibitem[Franks et~al.(2020)Franks, Airoldi, and Rubin]{franks2020nonstandard}
Alexander~M Franks, Edoardo~M Airoldi, and Donald~B Rubin.
\newblock Nonstandard conditionally specified models for nonignorable missing
  data.
\newblock \emph{Proceedings of the National Academy of Sciences}, 117\penalty0
  (32):\penalty0 19045--19053, 2020.

\bibitem[Ganguli et~al.(1998)Ganguli, Lytle, Reynolds, and
  Dodge]{ganguli1998random}
Mary Ganguli, Mary~E Lytle, Maureen~D Reynolds, and Hiroko~H Dodge.
\newblock Random versus volunteer selection for a community-based study.
\newblock \emph{The Journals of Gerontology Series A: Biological Sciences and
  Medical Sciences}, 53\penalty0 (1):\penalty0 M39--M46, 1998.

\bibitem[Garrett and Neufeld(2009)]{garrett2009invalid}
Brandon~L Garrett and Peter~J Neufeld.
\newblock Invalid forensic science testimony and wrongful convictions.
\newblock \emph{Virginia Law Review}, pages 1--97, 2009.

\bibitem[Groves(2006)]{groves2006nonresponse}
Robert~M Groves.
\newblock Nonresponse rates and nonresponse bias in household surveys.
\newblock \emph{Public opinion quarterly}, 70\penalty0 (5):\penalty0 646--675,
  2006.

\bibitem[Guttman et~al.(2022)Guttman, Laamanen, Russell, Atha, and
  Darnell]{guttman2022results}
Barbara Guttman, Mary~T Laamanen, Craig Russell, Chris Atha, and James Darnell.
\newblock Results from a black-box study for digital forensic examiners.
\newblock 2022.

\bibitem[Hicklin et~al.(2021)Hicklin, Winer, Kish, Parks, Chapman, Dunagan,
  Richetelli, Epstein, Ausdemore, and Busey]{hicklin2021accuracy}
R~Austin Hicklin, Kevin~R Winer, Paul~E Kish, Connie~L Parks, William Chapman,
  Kensley Dunagan, Nicole Richetelli, Eric~G Epstein, Madeline~A Ausdemore, and
  Thomas~A Busey.
\newblock Accuracy and reproducibility of conclusions by forensic bloodstain
  pattern analysts.
\newblock \emph{Forensic Science International}, page 110856, 2021.

\bibitem[Hicklin et~al.(2022)Hicklin, Eisenhart, Richetelli, Miller, Belcastro,
  Burkes, Parks, Smith, Buscaglia, Peters, et~al.]{hicklin2022accuracy}
R~Austin Hicklin, Linda Eisenhart, Nicole Richetelli, Meredith~D Miller, Peter
  Belcastro, Ted~M Burkes, Connie~L Parks, Michael~A Smith, JoAnn Buscaglia,
  Eugene~M Peters, et~al.
\newblock Accuracy and reliability of forensic handwriting comparisons.
\newblock \emph{Proceedings of the National Academy of Sciences}, 119\penalty0
  (32):\penalty0 e2119944119, 2022.

\bibitem[Hofmann et~al.(2021)Hofmann, Vanderplas, and
  Carriquiry]{inconclusives}
Heike Hofmann, Susan Vanderplas, and Alicia Carriquiry.
\newblock {Treatment of inconclusives in the AFTE range of conclusions}.
\newblock \emph{Law, Probability and Risk}, 19\penalty0 (3--4):\penalty0
  317--364, 05 2021.
\newblock ISSN 1470-8396.
\newblock \doi{10.1093/lpr/mgab002}.
\newblock URL \url{https://doi.org/10.1093/lpr/mgab002}.

\bibitem[Jakobsen et~al.(2017)Jakobsen, Gluud, Wetterslev, and
  Winkel]{jakobsen2017and}
Janus~Christian Jakobsen, Christian Gluud, J{\o}rn Wetterslev, and Per Winkel.
\newblock When and how should multiple imputation be used for handling missing
  data in randomised clinical trials--a practical guide with flowcharts.
\newblock \emph{{BMC} Medical Research Methodology}, 17\penalty0 (1):\penalty0
  1--10, 2017.

\bibitem[Jordan et~al.(2013)Jordan, Watkins, Storey, Allen, Brooks, Garaiova,
  Heaven, Jones, Plummer, Russell, et~al.]{jordan2013volunteer}
Sue Jordan, Alan Watkins, Mel Storey, Steven~J Allen, Caroline~J Brooks, Iveta
  Garaiova, Martin~L Heaven, Ruth Jones, Sue~F Plummer, Ian~T Russell, et~al.
\newblock Volunteer bias in recruitment, retention, and blood sample donation
  in a randomised controlled trial involving mothers and their children at six
  months and two years: a longitudinal analysis.
\newblock \emph{PLoS One}, 8\penalty0 (7):\penalty0 e67912, 2013.

\bibitem[Kerkhoff et~al.(2018)Kerkhoff, Stoel, Mattijssen, Berger, Didden, and
  Kerstholt]{kerkhoff2018part}
W~Kerkhoff, RD~Stoel, EJAT Mattijssen, CEH Berger, FW~Didden, and JH~Kerstholt.
\newblock A part-declared blind testing program in firearms examination.
\newblock \emph{Science \& Justice}, 58\penalty0 (4):\penalty0 258--263, 2018.

\bibitem[Khan and Carriquiry(2023)]{khan2023hierarchical}
Kori Khan and Alicia Carriquiry.
\newblock Hierarchical bayesian non-response models for error rates in forensic
  black-box studies.
\newblock \emph{Philosophical Transactions of the Royal Society A},
  381\penalty0 (2247):\penalty0 20220157, 2023.

\bibitem[Kim and Shao(2014)]{kim2014statistical}
Jae~Kwang Kim and Jun Shao.
\newblock \emph{Statistical methods for handling incomplete data}.
\newblock Chapman and Hall/CRC, 2014.

\bibitem[{Klees Expert Trial Transcript}(2022)]{CTS_testimony}
{Klees Expert Trial Transcript}.
\newblock {Proceedings Transcript of Daubert Hearing commencing on Wednesday,
  April 3, 2002 in United States v. Minerd, 2002 WL 32995663 (W.D.Pa.)}.
\newblock 2022.

\bibitem[{Kumho Tire Co. v. Carmichael}()]{Kumho}
{Kumho Tire Co. v. Carmichael}.
\newblock {526 U.S. 137 (1999)}.

\bibitem[Levy and Lemeshow(2013)]{levy2013sampling}
Paul~S Levy and Stanley Lemeshow.
\newblock \emph{Sampling of populations: methods and applications}.
\newblock John Wiley \& Sons, 2013.

\bibitem[Little and Rubin(2019)]{little2019statistical}
Roderick~JA Little and Donald~B Rubin.
\newblock \emph{Statistical analysis with missing data}, volume 793.
\newblock John Wiley \& Sons, 2019.

\bibitem[Madley-Dowd et~al.(2019)Madley-Dowd, Hughes, Tilling, and
  Heron]{madley2019proportion}
Paul Madley-Dowd, Rachael Hughes, Kate Tilling, and Jon Heron.
\newblock The proportion of missing data should not be used to guide decisions
  on multiple imputation.
\newblock \emph{Journal of clinical epidemiology}, 110:\penalty0 63--73, 2019.

\bibitem[Mislevy and Wu(1996)]{mislevy1996missing}
Robert~J Mislevy and Pao-Kuei Wu.
\newblock Missing responses and irt ability estimation: Omits, choice, time
  limits, and adaptive testing.
\newblock \emph{ETS Research Report Series}, 1996\penalty0 (2):\penalty0 i--36,
  1996.

\bibitem[Monson et~al.(2022)Monson, Smith, and Bajic]{monson2022planning}
Keith~L Monson, Erich~D Smith, and Stanley~J Bajic.
\newblock Planning, design and logistics of a decision analysis study: The
  fbi/ames study involving forensic firearms examiners.
\newblock \emph{Forensic Science International: Synergy}, 4:\penalty0 100221,
  2022.

\bibitem[Monson et~al.(2023)Monson, Smith, and Peters]{monson2023accuracy}
Keith~L Monson, Erich~D Smith, and Eugene~M Peters.
\newblock Accuracy of comparison decisions by forensic firearms examiners.
\newblock \emph{Journal of Forensic Sciences}, 68\penalty0 (1):\penalty0
  86--100, 2023.

\bibitem[{National Research Council (U.S.)}(2009)]{nrc}
{National Research Council (U.S.)}, editor.
\newblock \emph{Strengthening forensic science in the {United} {States}: {A}
  path forward}.
\newblock National Academies Press, Washington, D.C, 2009.
\newblock ISBN 978-0-309-13135-3 978-0-309-13131-5.

\bibitem[Pedersen et~al.(2017)Pedersen, Mikkelsen, Cronin-Fenton, Kristensen,
  Pham, Pedersen, and Petersen]{pedersen2017missing}
Alma~B Pedersen, Ellen~M Mikkelsen, Deirdre Cronin-Fenton, Nickolaj~R
  Kristensen, Tra~My Pham, Lars Pedersen, and Irene Petersen.
\newblock Missing data and multiple imputation in clinical epidemiological
  research.
\newblock \emph{Clinical epidemiology}, 9:\penalty0 157, 2017.

\bibitem[Pohl et~al.(2014)Pohl, Gr{\"a}fe, and Rose]{pohl2014dealing}
Steffi Pohl, Linda Gr{\"a}fe, and Norman Rose.
\newblock Dealing with omitted and not-reached items in competence tests:
  Evaluating approaches accounting for missing responses in item response
  theory models.
\newblock \emph{Educational and Psychological Measurement}, 74\penalty0
  (3):\penalty0 423--452, 2014.

\bibitem[{President's Council of Advisors on Science and
  Technology}(2016)]{pcast}
{President's Council of Advisors on Science and Technology}.
\newblock Forensic {Science} in {Criminal} {Courts}: {Ensuring} {Scientific}
  {Validity} of {Feature} {Comparison} {Methods}.
\newblock Technical report, September 2016.
\newblock URL
  \url{https://obamawhitehouse.archives.gov/sites/default/files/microsites/ostp/PCAST/pcast_forensic_science_report_final.pdf}.

\bibitem[Richetelli et~al.(2020)Richetelli, Hammer, and
  Speir]{richetelli2020forensic}
Nicole Richetelli, Lesley Hammer, and Jacqueline~A Speir.
\newblock Forensic footwear reliability: Part iii—positive predictive value,
  error rates, and inter-rater reliability.
\newblock \emph{Journal of Forensic Sciences}, 65\penalty0 (6):\penalty0
  1883--1893, 2020.

\bibitem[Riddles et~al.(2016)Riddles, Kim, and Im]{riddles2016propensity}
Minsun~Kim Riddles, Jae~Kwang Kim, and Jongho Im.
\newblock A propensity-score-adjustment method for nonignorable nonresponse.
\newblock \emph{Journal of Survey Statistics and Methodology}, 4\penalty0
  (2):\penalty0 215--245, 2016.

\bibitem[Rubin(1976)]{rubin1976inference}
Donald~B Rubin.
\newblock Inference and missing data.
\newblock \emph{Biometrika}, 63\penalty0 (3):\penalty0 581--592, 1976.

\bibitem[Schafer(1999)]{schafer1999multiple}
Joseph~L Schafer.
\newblock Multiple imputation: a primer.
\newblock \emph{Statistical methods in medical research}, 8\penalty0
  (1):\penalty0 3--15, 1999.

\bibitem[Schafer and Graham(2002)]{schafer2002missing}
Joseph~L Schafer and John~W Graham.
\newblock Missing data: our view of the state of the art.
\newblock \emph{Psychological methods}, 7\penalty0 (2):\penalty0 147, 2002.

\bibitem[Smith(2021)]{smith2021beretta}
Jaimie~A Smith.
\newblock Beretta barrel fired bullet validation study.
\newblock \emph{Journal of Forensic Sciences}, 66\penalty0 (2):\penalty0
  547--556, 2021.

\bibitem[Smith et~al.(2016)Smith, Andrew~Smith, and
  Snipes]{smith2016validation}
Tasha~P Smith, G~Andrew~Smith, and Jeffrey~B Snipes.
\newblock A validation study of bullet and cartridge case comparisons using
  samples representative of actual casework.
\newblock \emph{Journal of Forensic Sciences}, 61\penalty0 (4):\penalty0
  939--946, 2016.

\bibitem[Smith(1983)]{smith1983validity}
{TMF} Smith.
\newblock On the validity of inferences from non-random samples.
\newblock \emph{Journal of the Royal Statistical Society: Series A (General)},
  146\penalty0 (4):\penalty0 394--403, 1983.

\bibitem[{State v. Moore}()]{Moore}
{State v. Moore}.
\newblock {122 N.J. 420 (1991)}.

\bibitem[Strassberg and Lowe(1995)]{strassberg1995volunteer}
Donald~S Strassberg and Kristi Lowe.
\newblock Volunteer bias in sexuality research.
\newblock \emph{Archives of sexual behavior}, 24:\penalty0 369--382, 1995.

\bibitem[Taylor et~al.(2009)Taylor, Cahn-Weiner, and
  Garcia]{taylor2009examination}
Alex~M Taylor, Deborah~A Cahn-Weiner, and Paul~A Garcia.
\newblock Examination of volunteer bias in research involving patients
  diagnosed with psychogenic nonepileptic seizures.
\newblock \emph{Epilepsy \& Behavior}, 15\penalty0 (4):\penalty0 524--526,
  2009.

\bibitem[Ulery et~al.(2011)Ulery, Hicklin, Buscaglia, and
  Roberts]{ulery2011accuracy}
Bradford~T Ulery, R~Austin Hicklin, JoAnn Buscaglia, and Maria~Antonia Roberts.
\newblock Accuracy and reliability of forensic latent fingerprint decisions.
\newblock \emph{Proceedings of the National Academy of Sciences}, 108\penalty0
  (19):\penalty0 7733--7738, 2011.

\bibitem[{United States v. Cloud}()]{Cloud}
{United States v. Cloud}.
\newblock {No. 1:19-CR-02032-SMJ-1, 2021 WL 7184484 (E.D. Wash. Dec. 17,
  2021)}.

\bibitem[{United States v. Shipp}()]{Shipp}
{United States v. Shipp}.
\newblock {422 F.Supp.3d 762 (2019)}.

\bibitem[Wasserstein and Lazar(2016)]{wasserstein2016asa}
Ronald~L Wasserstein and Nicole~A Lazar.
\newblock The {ASA} statement on p-values: context, process, and purpose, 2016.

\bibitem[Weller and Morris(2020)]{weller2020commentary}
Todd~J Weller and Max~D Morris.
\newblock Commentary on: {I. Dror}, {N. Scurich} “{(Mis)} use of scientific
  measurements in forensic science” forensic science international: Synergy
  2020 https://doi. org/10.1016/j. fsisyn. 2020.08. 006.
\newblock \emph{Forensic Science International: Synergy}, 2:\penalty0 701,
  2020.

\bibitem[Yee(2015)]{yee2015book}
T.~W. Yee.
\newblock \emph{Vector Generalized Linear and Additive Models: With an
  Implementation in {R}}.
\newblock Springer, New York, NY, USA, 2015.

\bibitem[Yee(2023)]{yee2022package}
T.~W. Yee.
\newblock \emph{{VGAM}: Vector Generalized Linear and Additive Models}, 2023.
\newblock URL \url{https://CRAN.R-project.org/package=VGAM}.
\newblock R package version 1.1-8.

\bibitem[Yee and Wild(1996)]{yee1996}
T.~W. Yee and C.~J. Wild.
\newblock Vector generalized additive models.
\newblock \emph{Journal of the Royal Statistical Society, Series B,
  Methodological}, 58:\penalty0 481--493, 1996.

\bibitem[Yee(2010)]{yee2010vgam}
Thomas~W Yee.
\newblock The {VGAM} package for categorical data analysis.
\newblock \emph{Journal of Statistical Software}, 32:\penalty0 1--34, 2010.

\end{thebibliography}

\end{document}


\maketitle

\section{Sampling Bias} \label{si:samplingbias}
\begin{sloppypar}
The CV search was conducted with Westlaw Edge on April 15, 2022. The search was initially restricted to Expert Materials. The Boolean search terms used were:
\texttt{adv: PRA(crim! \& fire-arm ballistic) \& fire-arm rifle hand-gun shot-gun pistol gun weapon tool-mark /5 identif! match!}. The results were then limited to only ``Expert Resumes.'' The initial search yielded 201 resumes. These resumes were manually reviewed, as described below.
\end{sloppypar}
\begin{sloppypar}
First, resumes were reviewed to determine whether the expert's resume indicated the individual was an expert concerning firearm and toolmark identification. Some phrases that were considered in favor of demonstrating relevance are ``IBIS'', ``toolmark examinations'', and ''ballistics comparison and identification.'' An expert can have multiple areas of expertise. For this step, acknowledgment of expertise in identifying the type of firearm itself, reconstructing crime scenes, or bullet wound examination were, by themselves, insufficient to render a resume relevant. After this review, there were a total of 131 relevant resumes.
\end{sloppypar}
Second, resumes were reviewed to determine whether the expert's resume stated the expert was currently a member of the Association of Firearm and Tool Mark Examiners (AFTE). For this analysis, there were two possible categories: 1) the expert was a current member of AFTE, and 2) the expert was not a current member of AFTE.

Third, resumes were reviewed to determine the current position held by the individual. We recorded whether the position was private or public. Some factors that were used to determine the coding for this category is that “L.L.C.” typically demonstrates a private work position, while anything that includes mentioning Federal or state entities typically demonstrates a public work position. If it was not immediately apparent that a position involved a public or private employer, we used Google searches. For this analysis, there were three possible categories: 1) the expert was privately employed, 2) the expert was publicly employed, and 3) the resume did not state employment.

After all 131 resumes were reviewed, we restricted our analysis to unique experts. There were 60 unique individuals represented in the 131 resumes. In \cref{tab:resume_counts}, we describe the breakdown of the number of resumes by unique expert. Five experts had multiple resumes which disagreed on whether the individual was an AFTE member or publicly employed. One expert had potentially testified both while employed by a private entity and a public entity. One expert had potentially testified with resumes that had both missing current employment and a private employer as current employment. Three experts had at least one resume that indicated the individual was a current AFTE member and at least one resume that indicated the member was not a current AFTE member.  

These duplicates were resolved in favor of the FBI/Ames study criteria when possible. For purposes of analysis, the expert who potentially testified while being both privately and publicly employed was treated as being publicly employed, and the three experts who potentially testified both while a member of AFTE and while not a member of AFTE were treated as AFTE members. The one expert with a resume missing employment was treated as privately employed.

\begin{center}
\begin{tabular}{| L{4cm} |   L{4cm}  |     }
		\hline
		\rowcolor{tuftsblue}
		\makecell{\textbf{\textcolor{white}{\# Resumes Present}}} & \makecell{\textbf{\textcolor{white}{\# Experts}} }   \\
		\hline
		1  &  42  \\
		\hline
		2 &   8 \\
		\hline
		3 & 1 \\
		\hline
		4 & 3 \\ 
		\hline
		6 & 3 \\ 
		\hline
		8 & 1 \\
		\hline
		9 & 1 \\
		\hline
		23 &1 \\
		\hline 
\end{tabular}
\captionof{table}{\# of CVs per expert witness \label{tab:resume_counts} }
\end{center}

\section{Unit and Item Nonresponse Rates } \label{app:unititemnonresponse}
Here, we give more details about the unit and item nonresponse in black-box studies. These results are summarized in \cref{tab:nonresponse}, with additional details following the table. Unless otherwise specified, these numbers are across all analysis types (i.e., all relevant measures for accuracy, repeatability, and reproducibility as applicable). Not this data was comprehensively compiled as of September 2022. We made subsequent updates to the FBI/Ames study when an additional publication was released in \citet{monson2023accuracy}.

\begin{center}
\begin{tabular}{| L{6cm} |   C{3cm}|  C{3cm}|   }
		\hline
		\rowcolor{tuftsblue}
	\tiny	\makecell{\textbf{\textcolor{white}{Study}}} & \tiny \makecell{\textbf{\textcolor{white}{Unit Nonresponse}} } & 
	\tiny	\makecell{\textbf{\textcolor{white}{Item Nonresponse }}}  \\ 
		\hline
	\small	\cite{ulery2011accuracy} (Prints) & 33.5\% & 0.2\%\\
		\hline
	\small	\cite{baldwin2014study} (Cartridges) & 23.2\% & .06\% \\ 
		\hline 
		\small \cite{smith2016validation} (Cartridges \& Bullets) & 27.7\% & Incalculable \\
		\hline
	\small	FBI/Ames Study (Bullets)  & $\geq$ 32.4\% &  35.6\%  \\
		\hline
	\small	FBI/Ames Study (Cartridges)  & $\geq$ 32.4\% &  35.1\%  \\
		\hline
		\small \cite{eldridge2021testing} (Palmar Prints) & 31.1\% & 27.5\% \\
		\hline
	\small	\cite{hicklin2021accuracy} (Blood Splatter) &   Incalculable &  31.5\% \\
	\hline
	\small \cite{richetelli2020forensic} (Footwear impression) & 33.0\% & 9.6\% \\ 
		\hline 
	\small \cite{smith2021beretta} (Bullets) & 32.7\% & Incalculable \\
	\hline
	\small \cite{guttman2022results} (Digital) & Incalculable & Inapplicable \\ \hline
	\small \cite{hicklin2022accuracy} (Handwriting) & Incalculable & 16.3\% \\
	\hline
\end{tabular}
\captionof{table}{Example of nonresponse rates in black-box studies  \label{tab:nonresponse} }
\end{center}

\begin{itemize}
\item \cite{ulery2011accuracy}: The numbers in \cref{tab:nonresponse} reflect only the accuracy stage. There were insufficient data to calculate the nonresponse rates for the repeatability stage. The authors report the analyses were based on 169 participants and that each examiner was initially assigned 100 comparisons. The authors state  66.5\% of the participants who indicated interest completed the test. They then state that 3 participants returned incomplete tests and were dropped from subsequent analyses. For this paper, we treat these participants' 300 responses as item nonresponse. The authors also report that 27 items did not have a response. Some of these may have been missing because the study administrators did not present them; however, we cannot know how many. Thus, we treat 327 of a potential 17,200 responses as item nonresponses.

\item \cite{baldwin2014study}: The authors report that 284 participants agreed to participate, and 218 actually participated. The 218 were each sent 15 comparisons, and 2 comparisons were not reported or left blank.

\item \cite{smith2016validation}: The authors reported that 47 test kits were sent out and 34 were returned. The authors excluded 3 test sets submitted from the analysis without providing a detailed explanation for why. We treat these as item nonresponse.  The authors did not provide consistent information for calculating the item nonresponse; it is at least 8.8\%.

\item FBI/Ames Study: \citet{monson2022planning} reports that 270 potential participants contacted the Ames group to complete the study. After the exclusion of FBI examiners, 256 examiners reportedly received a first round of test sets. Only 173 examiners returned evaluations. The authors state that additional examiners joined the study but do not provide the number of additional examiners. If additional members beyond the initial 256 agreed to participate in the study, the estimate of unit nonresponse in \cref{tab:nonresponse} (1 - 173/256)*100 = 32.4\% is a lower bound for the unit nonresponse. \citet{monson2022planning} and \citet{bajic2020validation} reference that each of the 173 examiners received (or was intended to receive) 6 test packages. Each package contained 30 comparison decisions: 15 cartridges and 15 bullets. Of the 31,1140 comparison decisions sent out, \citet{chumbley2021accuracy} reported the examiners returned 20,130 decisions (10,110 cartridges and 10,020 bullets). Note this study included accuracy, repeatability, and reproducibility rounds, and the reported nonresponse rates are aggregated across all types of analyses.

\item \citet{eldridge2021testing}: The authors reported that 328 participants agreed to participate, and 226 returned responses to test sets. Each participant was assigned 75 comparisons (for a total of 16,950 comparisons). There were 12,279 comparisons with some kind of response. To view the missingness in the most favorable light to the study, we include those with missing comparison decisions as a response as long as the examiner at least evaluated the image quality.  We note that 2,819 assigned comparisons lacked a comparison decision. More specifically, 2,406 of the 12,279 comparisons treated as observed were marked as being of no value for comparison. None of these 2,406 comparisons received a comparison decision. An additional 413 comparisons deemed to be of value for comparison did not have a recorded comparison decision. Only $12,279-2,406-413=9,460$ comparisons of the originally assigned 16,950 comparisons received a comparison decision.

\item \citet{hicklin2021accuracy}:  The authors do not report the number of participants who originally enrolled or indicated an interest in the study. The authors reported results were obtained for 81 participants. Each participant was provided with 150 comparisons. The authors excluded 6 participants from all analyses because these participants had responded to less than 3 comparisons.  The authors did not report the exact number of comparisons returned. The item nonresponse was calculated using the medians of categories provided in the Table S8 of the Supplementary Materials for \citet{hicklin2021accuracy}. More specifically, we assume $(6 \times ((1+3)/2)+((5+45)/2 \times 19)+7\times ((58+96)/2)+((125+149)/2)\times 4+150 \times 45 =  8324$ comparisons received a response out of the potential $150 \times 81 = 12,150 $ assigned comparisons.

\item \citet{richetelli2020forensic}: reported that 115 examiners were recruited, and 77 submitted results. Each examiner was assigned 12 comparisons. The authors excluded 7 of the examiners who submitted results because they either had not completed training or had not performed at least one footwear comparison. Because the 7 examiners were allowed to submit results before the decision to exclude them, this article treats the excluded comparisons as item nonresponse rather than unit nonresponse. The authors reported 835 responses to comparisons out of a total of 77*12=924 assigned comparisons.

\item \cite{smith2021beretta}: The author reported that 110 participants agreed to participate and that 74 submitted responses. He reported the number of comparisons distributed but not the total number responded to.  The author of this study removed errors because the errors were likely ``administrative'' in nature and ``would'' have been caught in the real world. It was not possible to accurately calculate the number of nonresponses because of this and insufficient details about the study design. 

\item \citet{guttman2022results}: did not report sufficient details to calculate either unit nonresponse. The authors did report some data about ``skipped'' questions. However, their estimates did not directly correspond to the error rates we have focused on here so the item nonresponse would have to be defined differently.

\item \cite{hicklin2022accuracy}: The authors did not report how many participants originally volunteered or the number excluded for not meeting inclusion criteria. They reported that 86 participants were each assigned 100 comparisons and that there were a total of 7,196 responses recorded. 
\end{itemize}

\section{Case Studies}

\subsection{FBI/Ames Study: Bullets} \label{si:amesfbi}

When this paper was originally written, none of the FBI/Ames study data had been released. While the paper was under review, \citet{monson2023accuracy} released partial data on the accuracy stage of the study. In this section, we first explain the simulation study prepared before the release of the partial data. We then explore the simulation studies' implications in light of the released partial data.

The authors of the FBI/Ames study have not released detailed information about the study design (this remains true as of May 2023, to the best of our knowledge). The clearest description appears in the abstract of an unpublished report that is not currently available online. In the abstract of \citet{bajic2020validation}, the authors report that the plan was for each examiner to receive two packets for each of the three rounds of this study.  A single sentence in \citet{monson2022planning} also references that each of the 173 participants received 6 packages, although it does not explicitly reference how many packets should have been received for just the accuracy stage.  Each packet consisted of 15 bullet comparisons \citep{monson2022planning,bajic2020validation,monson2023accuracy}. Thus, there were a total of 15,570 assigned bullet comparisons in the study. \citet{chumbley2021accuracy} reports 10,020 bullet comparison decisions recorded. The item nonresponse rate is 35.6\%. 

For just the accuracy round, if each examiner was intended to complete two test packets(i.e., $15 \times 2 = 30$ bullet comparisons), then $15 \times 2 \times 173=5,190$ comparisons should have been assigned. In total, \citet{bajic2020validation} reports that 4,320 responses of the 5,190 assigned accuracy comparisons were received. However, an additional 138 \citep{bajic2020validation} of these responses were dropped from accuracy estimates. The authors report that the dropped responses ``includes records for which an evaluation
was not coded or was recorded as Inconclusive without a level designation (A, B, or C), where multiple
levels were recorded, or for which the examiner indicated that the material was Unsuitable for
evaluation.''  Thus, across all comparisons in the accuracy round, there is effectively $1 - \frac{ 4320-138}{5190} =  0.19$, or 19\% item nonresponse. 

In our simulation study, we focus on the estimate of the false positives. However, at the moment, there is not enough information to accurately calculate the item nonresponse rate for just different source items. If missingness is non-ignorable, as the percentage of missing items increases, the bias of the estimate obtained from an analysis that does not account for missingness will increase. To view the impact of the missingness in the light most favorable to the current estimates, we use the original study design to ensure that the missingness for the different source items in the accuracy round is as small as possible.

Specifically, relying on \citet{bajic2020validation} and \citet{monson2022planning}, we assume that there were two test packages per examiner for the accuracy round (for a total of 346 packets). Each of the 173 examiners should have been assigned 30 bullet comparison items.  In our simulation study, we focus on the estimate of the false positives. \citet{bajic2020validation} reports that $2/3$rds of the comparisons should have been different source comparisons. We note the study design did not ensure that each examiner should have seen 2/3rds of his/her assigned comparisons being different source \citep[see,][pg. 23]{bajic2020validation}. At the time of writing, the authors have not released information about the number or kinds of assigned items per examiner. Thus, we assume that each of the 173 examiners would see ($30 \times 2/3$) 20 different source items for the accuracy round. \citet{chumbley2021accuracy} reports that 2,891 decisions were recorded. However, they chose to restrict their analysis to 2,842 cases. Therefore, we assume that approximately 82.1\% (2842/3460) of the different source comparisons had a recorded response for the purposes of analysis. In other words, four our simulations, the item nonresponse is estimated to be $1 - .821 = .179$, or $17.9\%$.  

For the simulation study, we ensure that there is always approximately 17.9\% item nonresponse across all the examiners. To simulate our data, we let $Y_{ij}$ be an indicator of whether individual $i$ makes an error on item $j$. Let $M_{ij}$ be an indicator of whether individual $i$'s response to item $j$ is missing.  We assume that there are $173$ participants who are each given $20$ test sets. We generate the data such that there are approximately $17.9\%$ missing test sets in the following way: 
\begin{eqnarray*}
	P(Y_{ij} = 1) \overset{ind}{\sim} \textrm{Bern} \left(p_i \right), ~ i = 1, \ldots, 173, j= 1, \ldots, 20 \\ 
	P(M_{ij}= 1 | Y_{ij} = 1 ) \overset{ind}{\sim} \textrm{Bern} \left( \pi_i  \right)~ i = 1, \ldots, 173, j= 1, \ldots, 20  \\ 
	P(M_{ij}  = 1 | Y_{ij} = 0 ) ) \overset{ind}{\sim} \textrm{Bern} \left( \theta_i \right) ~ i = 1, \ldots, 173, j= 1, \ldots, 20 
\end{eqnarray*}   

For the first 163 individuals, $p_i$ is randomly generated from a uniform distribution on $[0,.007]$. For the other 10 individuals, $p_i$ is randomly generated from a uniform distribution on $[.55,.6]$.

We let $\pi_i = \pi$ be constant across individuals. For this set of simulations, we allow $\pi$ to vary between $0$ and $1$ at 101 equally spaced values. For each value of $\pi$, we simulated responses for all 173 examiners on 20 items. To ensure that approximately 17.9\% of the responses were missing, $\theta_i$  was chosen as a function of $\pi$ to ensure the appropriate percentage of missingness. There are many ways to choose $\theta_i$ to ensure approximately the appropriate missingness. 

For these simulations, the following method was used. We observe that Group A (the 163 low error rate examiners) and Group B (the 10 higher error rate examiners) account for approximately 94.2\% and 5.8\% of the 173 examiners, respectively. We use the fact that if $m_A$ and $m_B$ are the proportion of missing responses for Group A and B, respectively, then $.942 m_A + .058 m_B$ will be the overall missing rate for the whole study. 

We begin by assuming that each member of Group B responds to approximately 12 of the 20 items assigned to him/her (i.e., that .4 of the items were missing). Given $p_i$ and $\pi$ for a member of Group B, we let:
\[ \theta_i = \frac{.4  - p_i \pi}{1 - p_i} \] 

For each value of $\pi$, after we simulate all the data for Group B, we calculate the actual missingness for the simulated data $\hat{m}_B$. Then, we define $\hat{m}_A = \frac{.179 - .058 \hat{m}_B}{ .942}$, and for each member of Group A, we let: 
\[ \theta_i = \frac{\hat{m}_A  - p_i \pi}{1 - p_i} \].

The choices of $\theta_i$ need not theoretically be bound between 0 and 1. In these simulations, they were for all cases. The use of absolute values and/or a secondary choice of constant rejection could have been used if necessary. 

\begin{figure}
\begin{center}
\begin{minipage}{.6\textwidth}
    \includegraphics[width=.8\linewidth]{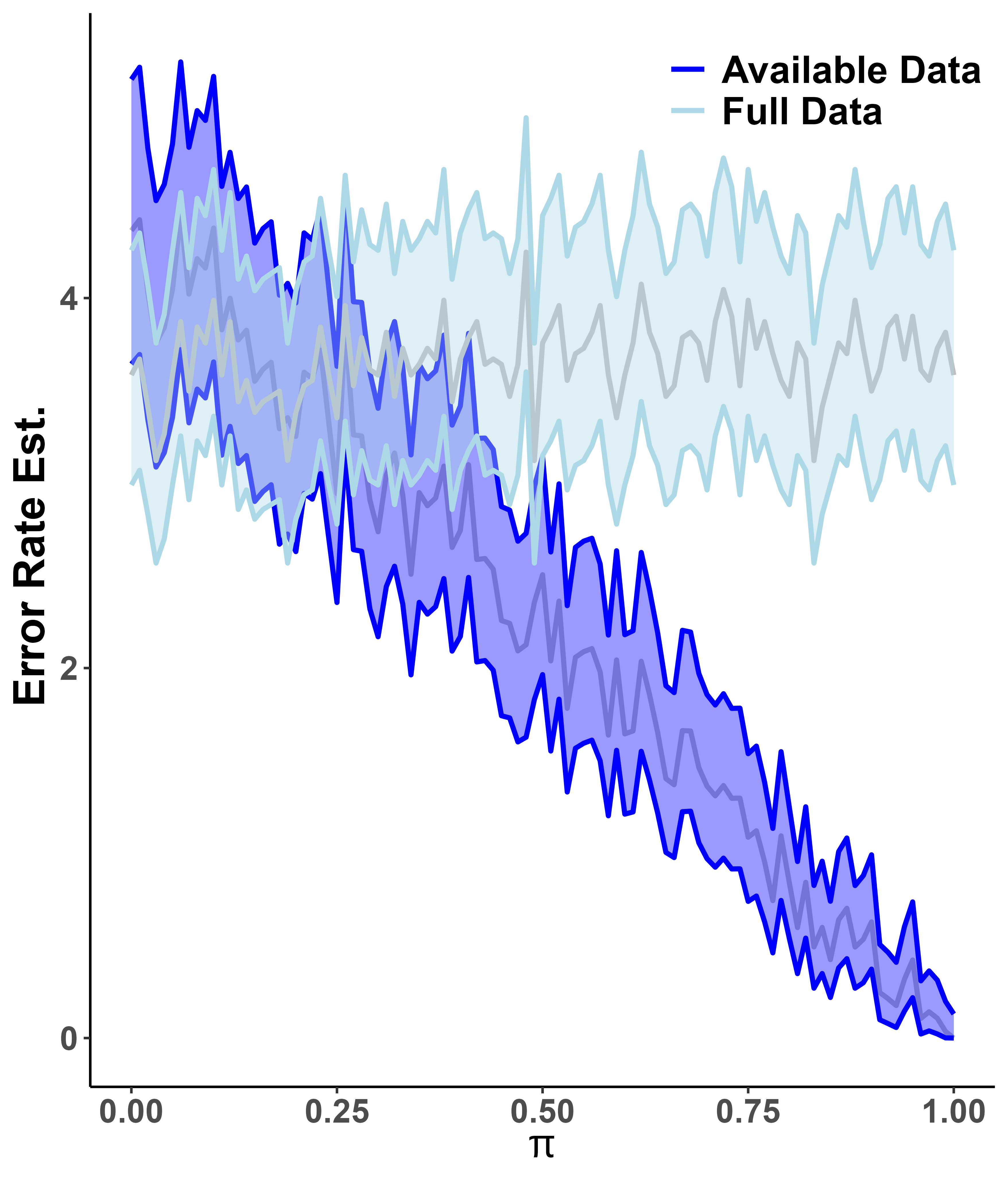}
    \caption{Clopper-Pearson Estimates and 95\% Confidence Intervals}
    \label{fig:comparing_est}
\end{minipage}
\end{center}
\end{figure}

In \cref{fig:comparing_est}, we demonstrate the 95\% Clopper Pearson confidence intervals for the false positive error rate. The dark blue represents an analysis of only the ``observed'' data and the light blue represents the analysis of the ``full'' data.

We now take a moment to compare these simulations to the partially released data in \citet{monson2023accuracy}. The released data included only the 4,320 items the 173 examiners responded to in the accuracy round. Importantly, they still did not provide sufficient information to calculate the item nonresponse for the different source and same source comparisons. This is because the authors did not release any information about the assigned items for each examiner. However, the released data do allow us to update the item nonresponse estimates for the accuracy range. Recall, for the accuracy round, \citet{bajic2020validation} states that each examiner was intended to complete two test packets(i.e., $15 \times 2 = 30$ bullet comparisons); however, data released in \citet{monson2023accuracy} showed one examiner completing 45 bullet comparisons in the accuracy stage. Thus, instead of the prior planned 5,190 accuracy comparisons, it seems that were at least ($172\times2 \times 15 + 3 \times 15)= $ 5,250 assigned accuracy comparisons. Of the 4,320 responses, the authors in \citep{monson2023accuracy} report dropping 139 (as opposed to the 138 in \citep{bajic2020validation}) of the 4,320 responses from analyses.  More information was provided about these in \citep{monson2023accuracy}. The authors state that some of these represented an examiner's decision that the item was unsuitable for comparison. However, another 26 responses were dropped for ``other'' reasons.  We are most concerned that the 17.9\% item nonresponse used in our simulations is an overestimate, so we only treat the 26 as missing so that there were $4,320-26=4,294$ observed responses and an associated $\frac{4,294}{5,250}=.182$, or 18.2\%, item nonresponse rate for the accuracy stage.  

For the entirety of the accuracy stage (both same source and different source), 59 examiners (34\%) of the 173 examiners had 50\% item nonresponse rates, and the rest had 0\% nonresponse rates (one examiner had 45 items, which we count as 0\% response rate, here). The number of different source items per examiner varied quite a bit: the first quartile was 11, and the third quartile was 21. The proportion of different source items per examiner was less variable The first quartile is .60, and the third quartile is .73. In any case, absent more explicit information about the number of different source items assigned to each examiner, we continue to rely on our previous estimates for the item nonresponse of the different source comparisons, as this is still lower than the overall item nonresponse in the accuracy stage. 

\section{Data and Code Availability}
All code used to conduct the analyses is available as part of the supplemental material. All data original to this study are also available. This includes the measures collected on the expert CVs considered in section 2 and the coding manual used to analyze them. The original CVs were accessed through a commercial resource with restrictions on public sharing. However, they are available for review upon request to \texttt{kkhan@iastate.edu}. The data from the EDC study is available as described in \citet{eldridge2021testing}, and the data from the FBI/Ames study is available as described in \citet{monson2023accuracy}.

\newpage


\bibliographystyle{plainnat}
\bibliography{references}